\documentclass{article}
\usepackage{algpseudocode}
\usepackage{algorithm}
\usepackage{graphicx}
\usepackage[utf8]{inputenc}
\usepackage{amsmath}
\usepackage{multirow}
\usepackage[official]{eurosym}
\usepackage{xcolor}

\usepackage{parskip}
\usepackage{subfigure}
\usepackage{caption}
\usepackage{wrapfig}
\usepackage{authblk}

\title{Rolling intrinsic for battery valuation in day-ahead and intraday markets}
\author[1]{Daniel Oeltz}
\author[2]{Tobias Pfingsten}
\affil[1]{\small Fraunhofer SCAI, Computational Finance}
\affil[2]{\small RIVACON GmbH}
\date{October 2025}

\begin{document}
\newcommand{\pnlunit}{\euro{}/day}

\maketitle
\begin{abstract}
    Battery Energy Storage Systems (BESS) are a cornerstone of the energy transition, as their ability to shift electricity across time enables both grid stability and the integration of renewable generation. This paper investigates the profitability of different market bidding strategies for BESS in the Central European wholesale power market, focusing on the day-ahead auction and intraday trading at EPEX Spot. We employ the rolling intrinsic approach as a realistic trading strategy for continuous intraday markets, explicitly incorporating bid--ask spreads to account for liquidity constraints. Our analysis shows that multi-market bidding strategies consistently outperform single-market participation.  Furthermore, we demonstrate that maximum cycle limits significantly affect profitability, indicating that more flexible strategies which relax daily cycling constraints while respecting annual limits can unlock additional value. 
\end{abstract}
\section{Introduction}
Battery Energy Storage Systems (BESS) are widely recognized as a cornerstone technology for enabling the energy transition. Their ability to decouple electricity generation and consumption in time makes them uniquely suited to address the intermittency of renewable energy sources such as wind and solar, thereby enhancing system stability, reliability, and efficiency. Beyond providing critical flexibility to balance supply and demand, BESS can support grid resilience, reduce curtailment of renewable generation, and defer costly network reinforcements. As decarbonization accelerates and variable renewables form the backbone of the electricity mix in Europe, the economic viability of BESS hinges on their ability to capture revenues from short-term power markets where volatility and arbitrage opportunities are most pronounced.

In recent years, the literature has devoted significant attention to the valuation of BESS in liberalized electricity markets. Early studies have focused on isolated markets, most prominently the day-ahead (DA) auction, where optimization-based bidding strategies rely on price forecasts and technical constraints to maximize arbitrage revenues \cite{Chen2016, MohsenianRad2016, Pozo2022}. Parallel strands of work have examined the continuous intraday (IDC) market (see \cite{Hornek2025} and references therein), highlighting its growing importance due to forecast errors from renewable generation and the resulting short-term imbalances. A range of methods have been explored, from mixed-integer linear programming to dynamic programming and reinforcement learning, with varying levels of abstraction regarding the complexity of market mechanisms, see \cite{Song2024} for an overview. However, simplifications—such as reducing the IDC to a small set of auctions or relying on index prices—have often underestimated the true potential of intraday trading. Recent studies show that rolling intrinsic strategies leveraging fine-grained transaction data more accurately capture arbitrage opportunities in the IDC market \cite{Semmelmann2024, Miskiw2025}. Schaurecker \cite{Schaurecker2025} further discusses these strategies in the context of high-frequency trading.

A second important strand of research investigates multi-market bidding, where BESS owners participate in both DA and IDC markets. Coordinated strategies that anticipate intraday conditions at the time of DA bidding have been shown to increase profitability, particularly for flexible storage assets \cite{Loehndorf2023}.  Recent work further emphasizes the practical importance of designing models and strategies that reflect real-world trading processes, including liquidity constraints and bid–ask spreads, to provide actionable insights for investors, operators, and policymakers \cite{Miskiw2025, Loehndorf2023}. 

Building on this literature, the present paper investigates the profitability of different trading strategies for BESS in the European wholesale power market, focusing on the EPEX intraday and day-ahead markets. We develop and evaluate multi-market bidding strategies based on the rolling intrinsic approach, while explicitly accounting for bid–ask spreads as a proxy for market liquidity and exploring the robustness of strategies with respect to the granularity of trading discretization. In doing so, we provide a systematic analysis of how operational strategies influence the profitability of BESS in practice.

The remainder of this paper is structured as follows: The next section describes the relevant market design and provides descriptive statistics on liquidity in the intraday market. Section three introduces the optimization framework, including the rolling intrinsic strategy and the derivation of bid and ask prices from transaction data. We present results in section four, where we address three key research questions: (i) the relative performance of different multi-market strategies, (ii) the sensitivity of profitability to liquidity via bid–ask spreads, and (iii) the robustness of rolling intrinsic trading with respect to aggregation of transaction data. In the last section we summarize the main insights and outline promising directions for further work, including extensions to other market segments and integration with stochastic forecasting approaches.

\section{BESS in Central European power markets}

In the following we provide a short introduction into the Central European power market as the main source of revenues for BESS in that area. For a more detailed introduction into markets and revenue streams see \cite{Riva2025}. 

The first distinguishing feature in the battery business case is whether it is placed in-front-of-the-meter or behind-the-meter. Behind-the-meter batteries are co-located with consumption and/or generation assets such as PV, their business case  typically being driven by maximizing consumption of locally produced electricity. We will focus on the in-front-of-the-meter business case, where BESS are directly participating in power wholesale markets.

\subsection{Market design}

\begin{figure}
\centering
    \includegraphics[width=1.0 \textwidth]{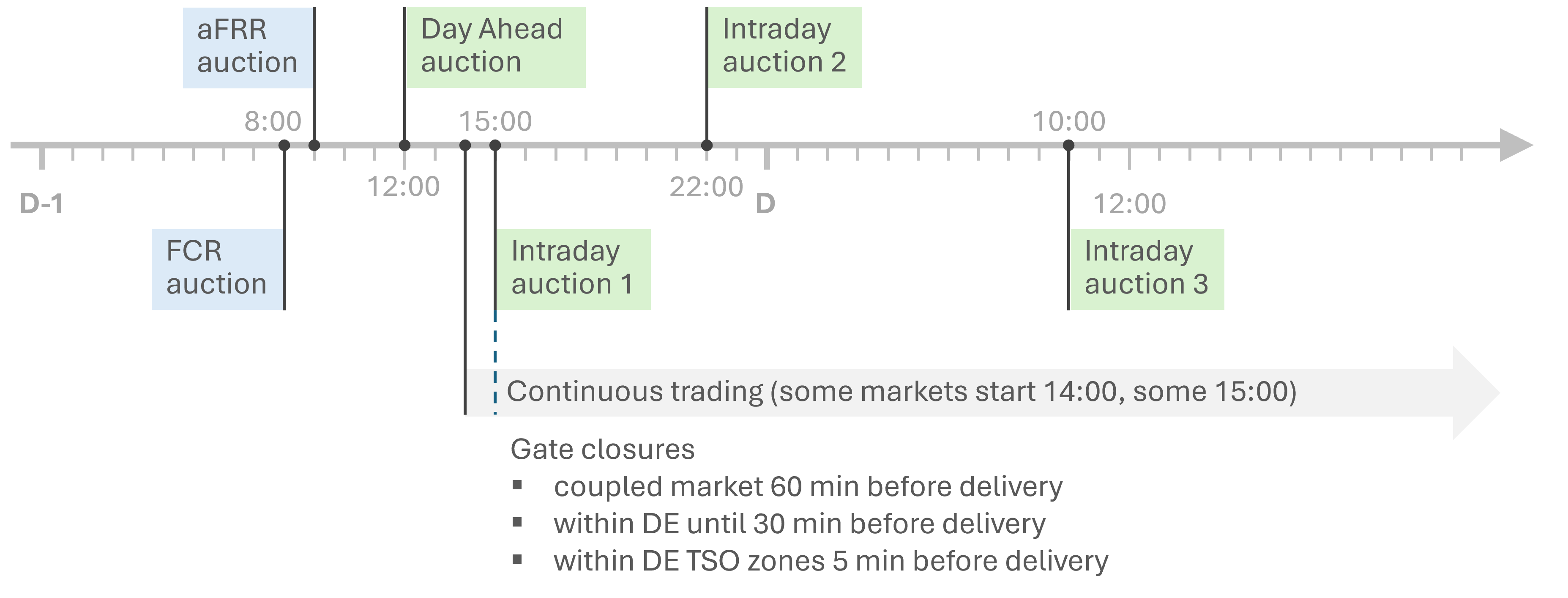}
\caption{Day-ahead and intraday markets as organized by EPEX Spot for Germany.}
\label{fig:timeline}
\end{figure}

Let us introduce the Central European power market in some detail to set the scene for how BESS capacity is monetized and what products are traded. There are several market places or auctions, that are organized on a daily basis. Figure~\ref{fig:timeline} provides an overview. 
Throughout this paper we refer to auctions and market places provided by EPEX Spot. However, note that there also other exchanges that provide similar markets places. Such are, for example, NordPool or EXAA.

\paragraph{Reserve Markets:} Auctions for Frequency Containment Reserve (FCR) or Frequency Restauration Reserve (FRR) take place in the morning the day before delivery.
Here, flexible capacity may be sold to the grid operator, which then utilizes the capacity to stabilize the grid. In case the BESS capacity is sold into reserve markets, it is no longer available to other purposes such as doing time-arbitrage in the intraday market. The Electricity Balancing Guideline provided by ENTSO-E~\cite{ENTSOE-network-code} provides a detailed introduction to reserve markets.

Reserve markets play a significant role for BESS. They are particularly suited for FCR with its high requirements on fast reaction times. As set out in \cite{Riva2025}, the decision whether to place BESS capacity into reserve markets or into day-ahead or intraday power markets for arbitrage, is highly relevant to maximize revenues. In this paper we concentrate on the latter to focus on a detailed analysis of intraday trading strategies.

\paragraph{Day-ahead (DAH) auction:}
The day-ahead auction at 12:00 (example EPEX for Germany) on the day before delivery is the main vehicle for consumers and generators to optimize their assets. Being an auction, market participants send in their bids, the exchange builds an order book and clears the market. Through Single Day-ahead Coupling (SDAC), the order books of all Central European exchanges\footnote{More specifically \textit{Nominated Electricity Market Operators} or short NEMOs.} are combined to create a single market up to limits given by cross-border transmission capacities. The pricing mechanism is \textit{pay as cleared}, meaning that the same price holds for all successful bids independently from their bidding price. Clearing is currently done on hourly, effectively creating hourly products. It is planned to change to a 15~min resolution in October 2025.

\paragraph{Block orders:}
A challenge for market participants and in particular for BESS optimization is to generate meaningful bids for the DAH auction -- since prices are available only after the auction has cleared. In order to design suitable orders, the operator will have to resort to a forecast to identify promising time-slots in which to charge and discharge the BESS. Only if those prove to coincide with the optimal time-slots of the auction result, the resulting revenue is maximized. 
However, suitable power price forecasts are available in the market, and clearing prices from other, earlier, auctions such as provided by EXAA can provide good estimates.

EPEX Spot (as other exchanges) provides the possibility to utilize "Block Order" to bid into the DAH auction. In addition, there are "Loop Order", specifically designed for storage assets. Assuming we have optimized the BESS against a price forecast, all hours of charging are collected in, say, block A, all hours of discharging in block B. The Loop Order instructs the exchange only to accept the complete order consisting of both blocks, if combined they exceed the required revenue.
This way, the operator can ensure to come out of the auction with obligations that may be fulfilled physically with the BESS asset. A case where only discharge bids are accepted, not the charge bids, will not occur and hence the risk of open positions is eliminated.

\paragraph{Intraday auctions:}
The DAH auction is the first opportunity for market participants to balance their portfolios or to optimize the utilization of their assets. However, as the time of actual delivery comes closer, weather forecasts (for PV and wind generation) or forecasts of plant availability or demand are improved -- and supply and demand need to be re-adjusted. 

There are three intraday auctions provided by EPEX Spot to do this, which are designed similarly to the DAH auction. They provide a good vehicle to place structured orders such as Loop Orders, however, liquidity is typically much smaller than ins the DAH auction. 
In 2024, traded volumes on EPEX for Germany were 291~TWh in the DAH auction, 91~TWh in the continuous intraday market compared to only 11~TWh in the intraday auction \cite{EPEX-Annual24}.

\paragraph{Intraday (ID) continuous trading:}

In ID continuous, power can be traded until 5~min before delivery. Until 60~min before delivery, order books are coupled across exchanges and country borders, below 30~min before delivery, order books are separated into the four TSO areas in Germany.  In contrast to DAH or ID auctions, orders are not cleared at a specific point in time -- and there is thus no "ID price" as there is as a result of the DAH auction. Instead, market participants continuously place orders in the exchanges' order books, which may or may not be cleared with other participants' orders.
As noted above, the traded volume (91~TWh in 2024 on EPEX) is significant as compared to an overall consumption of 465~TWh net consumption in Germany \cite{ENTSOE-factsheet}.

To provide an approximate picture of ID continuous market results, EPEX and other exchanges provide price indices. In the case of EPEX, those are ID1, ID3 and IDFull, which represent a weighted average across all trades in the last 1~or~3~hours before delivery and across all trades, respectively. TSOs utilize the AEP index, which averages across the last 500 MW traded for each 15 min product as a component in the calculation of the imbalance price~\cite{TSOsAEP}.

\begin{figure}
\centering
    \includegraphics[width=1.0 \textwidth]{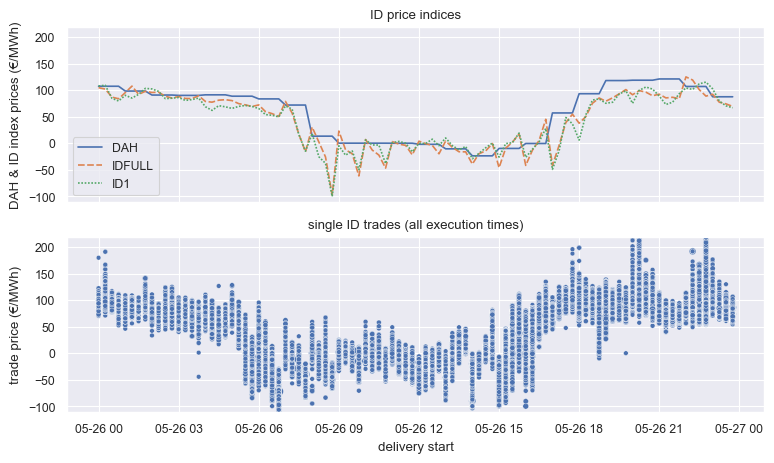}
\caption{DAH prices, intraday indices and trades on EPEX Spot. Trades beyond limits on the y-axis cut off for improved readability. 15~min products only, dot size reflects trade volume. Note that there are $>300.000$ ID trades in that timeframe.}
\label{fig:indices_trades}
\end{figure}

Figure~\ref{fig:indices_trades} shows a sample of these indices together with the corresponding intraday trades. Note that generally speaking, the ID1 index is more volatile than the IDFull and ID3 index, since it encompasses only the trades of the last hour before delivery. Indices as averages reflect only a small part of activity in ID continuous, as indicated by the large range of trade prices across execution times.

%%% INDEX figure.. als PDF zu groß, daher PNG

\paragraph{BESS optimization in ID continuous:}
The very nature of ID continuous makes battery optimization a challenging task, since there is no tradeable price curve available across all 15 min products for the coming hours. It is thus not possible to reduce BESS optimization to running an optimizer across the price curve to determine the optimal times to charge and discharge. 

A trading strategy is required to come from single trades and orders for specific products at a specific point in time to  optimized battery dispatch and revenue.
In section~\ref{sec:prob-formulation} of this paper we utilize a rolling intrinsic hedge approach that can actually be applied in real-life trading strategies.

Many authors utilize price indices such as the ID1 to approximate BESS revenues ex-post. The choice of the index, however, constitutes a strong assumption for revenues, as shown in figure \ref{fig:indices_trades}. The less time before delivery is included, the higher volatility generally is and the higher approximated revenues. In addition, bid/offer spreads are not reflected when using indices. This caveat should be borne in mind when interpreting results based on index prices. Section \ref{sec:results} presents index-based results and compares them with the rolling intrinsic approach.

%%%%%%%%%%%%%%%%%%%%%%%%%%%%%%%%%%%%%%%%%%%%%%
\subsection{Market Statistics}\label{sec:market-statistics}

As previously discussed, the intraday market is considerably more complex than auction-based trading. Section \ref{sec:results} shows that participation in continuous intraday trading can substantially increase profits. Thus, the intraday market may play a key role in the monetization of BESS in wholesale electricity markets. However, due to its structural complexity compared to the auction market, a fundamental understanding of its organization is essential. Therefore, we present and analyze basic market statistics in this section.

\begin{figure}
\centering
    \includegraphics[width=1.0 \textwidth]{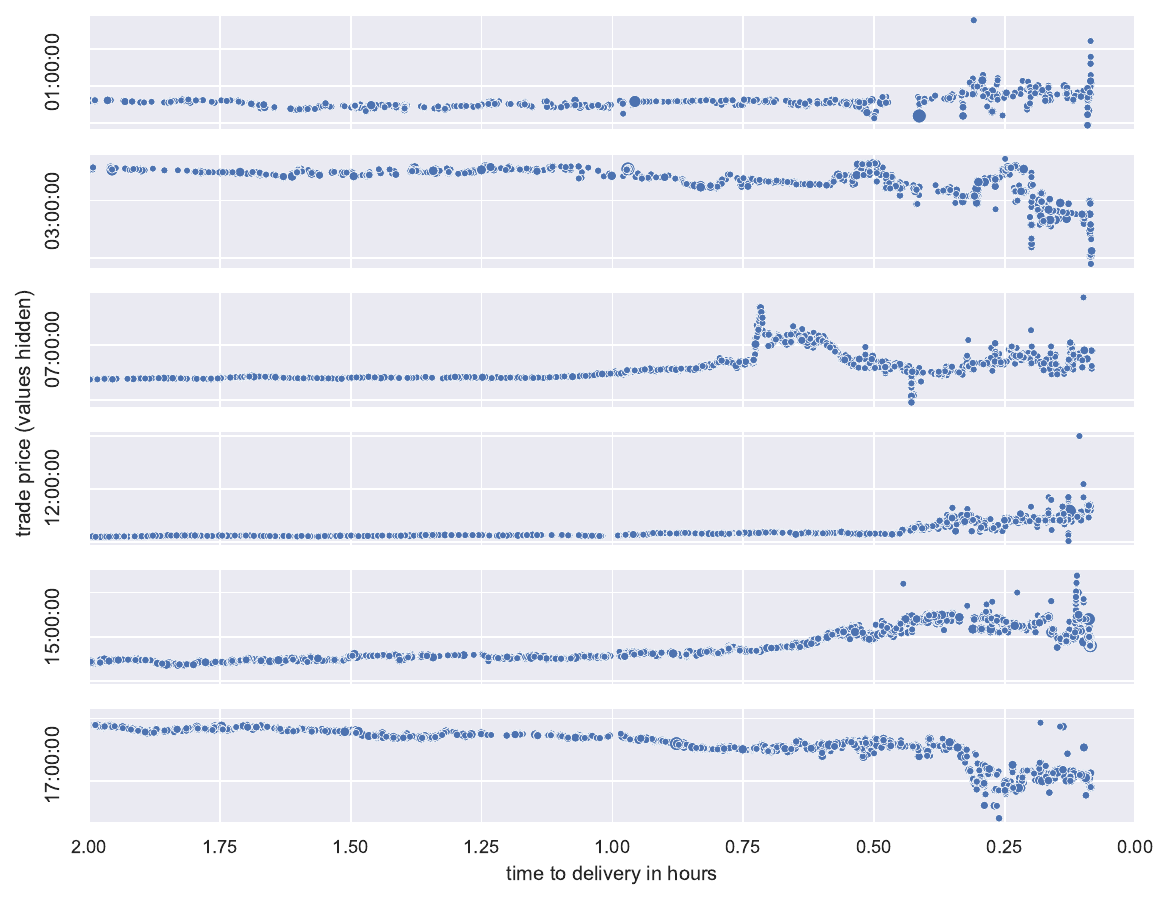}
\caption{Illustration: Trades for 15~min products on  24/03/30. Single graphs show trades for specific 15~min products, the x-axis shows the time to delivery in hours, dot size reflects trade volume.}
\label{fig:sample_prods}
\end{figure}

In Figure~\ref{fig:sample_prods} we illustrate the results for a specific day and selected 15~min products. Each dot represents a trade that has been conducted, its size representing the trade's volume. The y-axis for each sub-chart shows the trade price, the x-axis the time before delivery. Note that there continuously are trades being executed at varying prices. We can clearly see that trading activity increases with decreasing time to the delivery period.

\begin{figure}
\centering
  \subfigure[Volume share\label{fig:vol}]{\includegraphics[width=0.48\textwidth]{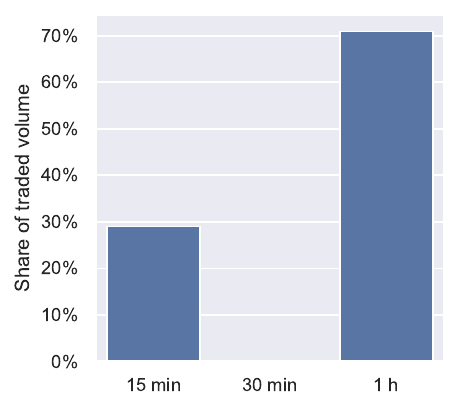}}
  \subfigure[No. of trades\label{fig:no}]{\includegraphics[width=0.48\textwidth]{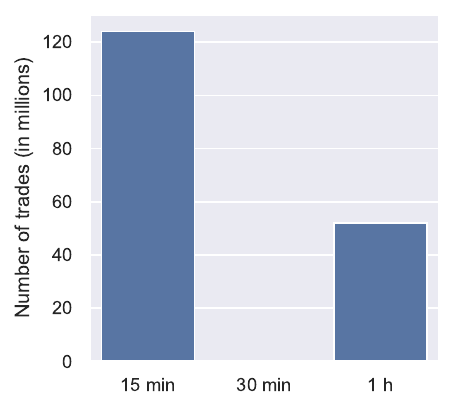}}
\caption{Traded volume and number of trades per product in 2024. User defined blocks are not considered.}
\label{fig:vol_prods}
\end{figure}

In IDC, quarters, half-quarters or hours of the corresponding delivery day are traded.  Figure~\ref{fig:vol_prods} shows traded volume (a) and the number of trades (b) per product. Note that 30~min are hardly traded, most volume coming from quarters and hours. While more volume is traded in hours, the largest share in terms of the number of trades comes from quarters\footnote{User defined blocks may also be traded, however, in 2024 corresponding volume was very low.}

% LIQUIDITY
\begin{figure}
\centering
  \subfigure[Cumulative volume by trading phase\label{fig:liq_phase}]{\includegraphics[width=1.0    \textwidth]{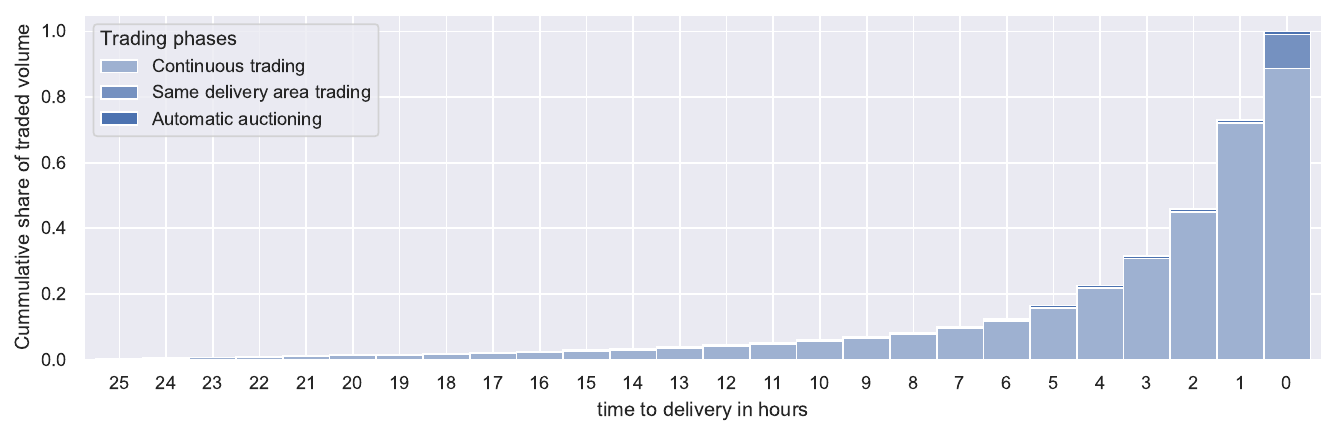}}
  \subfigure[Cumulative number of trades by product\label{fig:liq_prod}]{\includegraphics[width=1.0 \textwidth]{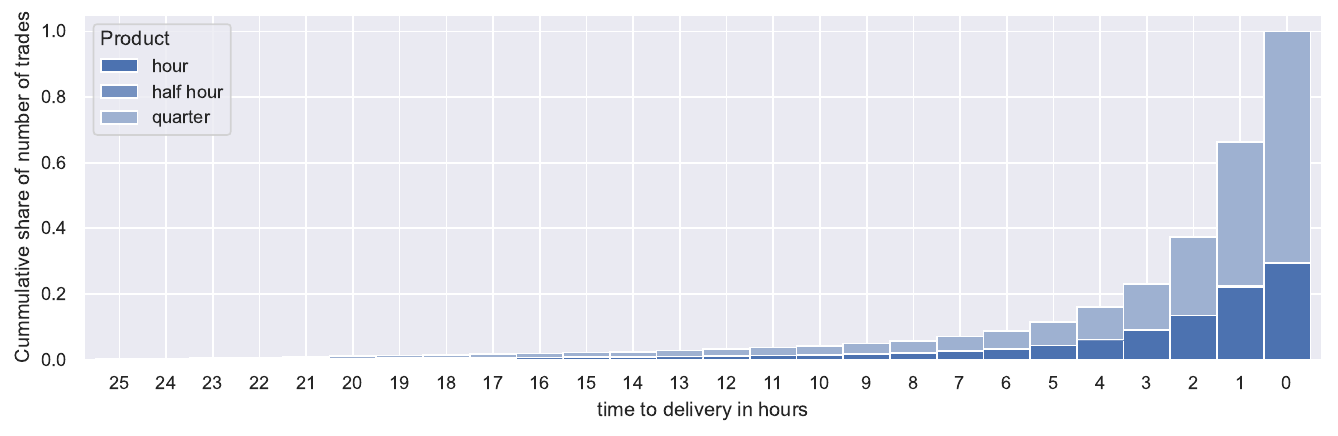}}
\caption{Liquidity in ID continuous against time to delivery. Liquidity is concentrated to the last hours before delivery, though markets open much earlier. Both, for hour and quarter products (b). The effect is more pronounced for the number of trades (b) than for volume (a). 2024 data.}
\label{fig:liquidity_overall}
\end{figure}

Figure~\ref{fig:liquidity_overall} shows that trading activity (i.e. liquidity) is concentrated to the last hours before delivery. For battery optimization this means, that liquidity further away is limited and there may be a high bid-ask spread. Note that in the last 30~min, order books decompose into delivery zones ("same delivery area trading" phase in figure~\ref{fig:liq_phase}). During continuous trading, exchanges may clear the market when cross-border capacity is released in automated auctioning ("automatic auctioning" in the figure, volume being comparably small). 
Figure~\ref{fig:liq_prod} shows liquidity in terms of the number of trades by products traded. We observe that concentration to the last hours before delivery is more pronounced than for volume, most likely due to "fine-tuning" of positions using many trades with smaller volume -- typically on quarter products.

% würde ich rausnehmen
%\begin{figure}
%\centering
%    \includegraphics[width=1.0 \textwidth]{figures/total_traded_volume.png}
%    \includegraphics[width=1.0 \textwidth]{figures/traded_volume_per_30min_before_delivery.png}
%\caption{Total trading volumes for each day.}
%\end{figure}

%\begin{figure}
%\centering
%    \includegraphics[width=1.0 \textwidth]{figures/traded_volume_per_trade.png}
%\caption{Average traded volume for each day.}
%\end{figure}

%\begin{figure}
%    \centering
%    \includegraphics[width=0.49\textwidth]{figures/total_num_trades_weekday.png}
%    \includegraphics[width=0.49\textwidth]{figures/total_traded_volume_month.png}
%    \caption{Average traded volumes per weekday (left, monday=0,...) and month (right).}
%    \label{fig:my_label}
%\end{figure}

\section{Problem Formulation}\label{sec:prob-formulation}
In this section we describe the underlying optimization problem as well as the derived rolling intrinsic approach that serves as the foundation of the trading strategies. 

\paragraph{Optimization Problem:}
A battery energy storage system (BESS) can be described by the following quantities: 
\begin{itemize}
    \item The maximal energy capacity $\overline{\mbox{SoC}}$ [MWh],
    \item the charge/discharge efficiencies $\eta^+$, $\eta^-$, 
    \item the maximum charge/discharge power $\bar{P}$ [MW],
    \item maximum number of charging cycles $N^{cycles}_{max}$ per day\footnote{Maximum cycle constraints are typically imposed by warranty agreements and are specified on an annual basis. However, to ensure that these constraints are not violated by trading strategies— which are often optimized over short horizons ranging from one to several days— it is practical to reformulate the annual limits as equivalent daily constraints, enabling more effective and explicit enforcement within the optimization framework.}.
\end{itemize}

Let us consider an equidistant time grid $\{t_i\}_{i=0}^N$ with step size 
$\Delta t = t_{i+1} - t_i$, where $\Delta t$ corresponds to the delivery 
periods of the traded products (one hour in the day-ahead market and 
15 minutes in the intraday market). Let $c_i$ and $d_i$ denote the charging 
and discharging power, respectively, applied during the interval 
$[t_i, t_{i+1})$. The state of charge (SoC) of the battery at time $t_i$ 
is denoted by $\mathrm{SoC}_i$, with initial level $\mathrm{SoC}_0$.

The battery operation is subject to the following constraints:
\begin{align}
\mathrm{SoC}_{i+1} = \mathrm{SoC}_i + \eta^+ c_i \Delta t 
    - \frac{1}{\eta^-} d_i \Delta t, \; 0\leq i \leq N-1, \label{eq:soc-dynamics} \\
0 \leq c_i, d_i \leq \bar{P},\;  0 \leq i \leq N, \label{eq:power-limits} \\
c_i \cdot (1-b_i) = 0, \quad d_i \cdot b_i = 0, 
    b_i \in \{0,1\}, \; 0\leq i\leq N, \label{eq:mutual-exclusion} \\
\sum_{i=0}^{N-1} c_i\Delta t \leq N^{\mathrm{cycles}}_{\max} \, \overline{\mathrm{SoC}}.
    \label{eq:cycle-limit}
\end{align}

Equation~\eqref{eq:soc-dynamics} defines the state-of-charge dynamics, 
accounting for charging efficiency $\eta^+$ and discharging efficiency 
$\eta^-$. Constraint~\eqref{eq:power-limits} limits charging and discharging 
power to the rated capacity $\bar{P}$. 
Constraint~\eqref{eq:mutual-exclusion} enforces that the battery cannot be 
charged and discharged simultaneously.\footnote{Without this restriction, 
simultaneous charging and discharging could become mathematically optimal 
at negative prices and round-trip efficiencies below one.} 
Finally,~\eqref{eq:cycle-limit} imposes a limit on the total energy throughput, 
which serves as a proxy for restricting the maximum number of cycles.

Since we restrict the analysis to a single day in accordance with the 
day-ahead and intraday markets, consistency across consecutive days is 
ensured by requiring the terminal SoC to equal the initial level:
\begin{equation}
    \mathrm{SoC}_N = \mathrm{SoC}_0.
    \label{eq:end-condition}
\end{equation}

Let us assume that for each time point $t_i$ we have a price $p_i^a$ to buy power (ask price) and a price to sell power $p_i^b$ (bid price), $p_i^b\leq p_i^a$. We can then determine the optimal dispatch by solving the  problem
\begin{equation}
\max_{c_i,d_i, b_i} \Delta t \sum_i p_i^b d_i - p_i^ac_i, \mbox{ s.t. (\ref{eq:soc-dynamics})-(\ref{eq:end-condition}) hold true.} \label{optim-problem}
\end{equation}
 
Note that we have to multiply (\ref{eq:soc-dynamics}) as well as  (\ref{optim-problem}) by $\Delta t$ to account for different delivery period lengths according to the traded product. While the day ahead auction still trades hourly products\footnote{As-of time of writing, it is planned to switch to quarter hourly products in October 2025} ($\Delta t = 1$), the products for the continuous trading cover quarter, half and full hourly delivery periods. We will use the products with quarter hourly delivery ($\Delta t=0.25$) due to their higher liquidity compared to the other products.

In both the rolling intrinsic approach and multi-market bidding, it is necessary to account for an initial state of charge and discharge already placed on the market, i.e. the current position, denoted by $\bar c_i$ and $\bar d_i$, respectively. These initial conditions must be incorporated into the formulation of the optimization problem in (\ref{optim-problem}). Therefore we define the residual quantities $c_i^r$ and $d_i^r$ that must be traded to achieve the new dispatch by 
\begin{align}
    c_i^r(c_i,d_i;\bar c_i,\bar d_i):= \max(c_i-\bar c_i,0)+\max(\bar d_i - d_i,0),\\
    d_i^r(c_i,d_i; \bar c_i, \bar d_i) :=\max(d_i-\bar d_i,0)+\max(\bar c_i - c_i,0).
\end{align}
\paragraph{Rolling Intrinsic:}
For the intraday continuous trading we apply a rolling intrinsic strategy similar to \cite{Semmelmann2024}. Here, we define a second time grid $\{t_j^T\}_{j=1}^M$ with time points where we might rebalance our positions. At each trading time point $t_j^T$, updated bid and ask prices, denoted by $p_{i,j}^b$ and $p_{i,j}^a$, are observed. The methodology for deriving these prices from EPEX market data will be addressed in a subsequent section. 
Note that whenever we step over a time point where delivery of a product starts, we realize delivery and the initial state of charge $SoC_0$ that is used in the next optimization must be modified accordingly. Moreover, this does also affect the max cycle constraint (\ref{eq:cycle-limit}) and we slightly modify this to account for charging in past time steps $t_i\leq t^T_j$ to 
\begin{equation}
    \hat c_j + \sum_i c_i  \leq N^{cycles}_{max}\overline{\mbox{SoC}}, \label{max-cycle-cond-ri}
\end{equation}
where $\hat c$ denotes the previous, realized total charged volume of the current day.

For each trading time point $t_j^T$ this leads to the problem
\begin{equation}\label{ri-optim}
    \max_{c_i,d_i, b_i} \Delta t \sum_{i: t_i<t^T_j} p_{i,j}^b d_i^r(c_i,d_i; \bar c_i, \bar d_i) - p_{i,j}^ac_i^r(c_i,d_i;\bar c_i,\bar d_i), 
\end{equation}
 s.t. (\ref{eq:soc-dynamics})-(\ref{eq:mutual-exclusion}), (\ref{eq:end-condition}) and (\ref{max-cycle-cond-ri}) hold true.
The overall algorithm is shown in \ref{ri-algorithm}.
\begin{algorithm}
\begin{algorithmic}
\State SoC = $\mathrm{SoC}_0$, 
\State value $\gets$ 0
\State $\bar c_i\gets 0$
\State $\bar d_i \gets 0$
\For{$k = 1, \dots, M$}
    \State Compute $c_i$, $d_i$, $b_i$ by solving (\ref{ri-optim}) at point $t^T_k$
    \State Update $\bar c_i$, $\bar d_i$ from solution of (\ref{ri-optim})
    \State value $\gets$ value + solution of (\ref{ri-optim})
    \State $\mathrm{SoC}_0 \gets \sum_{i: t_{k-1}\leq t_i\leq t_k}\eta^+c_i - d_i/\eta^-$
\EndFor
\end{algorithmic}
\caption{Rolling intrinsic algorithm for intraday continuous trading.}\label{ri-algorithm}
\end{algorithm}
\paragraph{Bid-Ask Prices:}
To apply the rolling intrinsic strategy, we need to  derive $p_{i,j}^b$ and $p_{i,j}^a$ at each trading time $t^T_j$ and for each product with delivery start at $t_i$ from the EPEX trade data. Let us denote the price of a trade at time $t$ for product with delivery start at $t_i$ by $p_{i,t}$. For $t^T_j$ we define the set of all prices in the preceding time bucket by $P_{i,j}:=\{ p_{i,t}\mid t^T_{j-1}<t\leq t^T_j\}$. 
We then define the bid price $p_{i,j}^b$ by
\begin{equation}\label{bid-price}
    p_{i,j}^b := \left\{\begin{array}{l} q_k(P_{i,t_j}) \mbox{ if } |P_{i,j}| \geq N,\\
    -4000 \mbox{ otherwise.}
    \end{array} \right. 
\end{equation}
Here, $q_k(P_{i,t_j}) $ denotes the empirical $k$-quantile for a fixed value of $0<k\leq |P_{i,t_j}| $ and $N$ a threshold parameter to define the minimum number of trades that must have been made in the bucket.
We apply the logic analogously for the ask prices and get
\begin{equation}\label{ask-price}
    p_{i,j}^a := \left\{\begin{array}{l} q_{-k}(P_{i,t_j}) \mbox{ if } |P_{i,j}| \geq N,\\
    4000 \mbox{ otherwise,}
    \end{array}\right.
\end{equation}
where $q_{-k}$ is the empirical $|P_{i,t_j}|-k$-quantile. 
If not stated otherwise, we set $N=10$ analogously to \cite{Semmelmann2024, Miskiw2025} and use the empirical $20\%$ quantile for bid prices and $80\%$ for ask prices respectively.

To solve the optimization problems presented above, we relied on the EAO software package \cite{EAOTechreport2021}, which builds upon the source code provided in \cite{EAO}.
\section{Results}\label{sec:results}

\begin{table}
    \centering
    \footnotesize
    \begin{tabular}{l|c|c|c}
        &  1h-battery & 2h-battery & 4h-battery\\\hline
        $\bar P$ & 2 & 1 & 0.5\\
        $\overline{\mbox{SoC}}$ & 2 & 2 & 2\\
         $\mbox{SoC}_0$ & 0.5       & 0.5        & 0.5\\
         $\mbox{SoC}_T$ & 0.5       & 0.5        & 0.5\\
         $N^{cycles}_{max}$ & 1     & 1          & 1\\
         $\eta^+$ & 97\% & 97\% &97\% \\
         $\eta^-$ & 98\% & 98\% & 98\%\\
    \end{tabular}
    \caption{Three different battery configurations.}
    \label{tab:battery_confiurations}
\end{table}

In this section, we present and analyze the performance of various multi-market bidding strategies, including the rolling intrinsic approach, using historical EPEX market data from June 14, 2024, to July 1, 2025.  We use three different battery configurations as shown in table \ref{tab:battery_confiurations}, where the batteries differ just by their charge and discharge capabilities, i.e. we have a one-hour battery (ignoring efficiencies fully charged after one hour), a two-hour battery and a four hour battery.

Our analysis is guided by the following key research questions:
\begin{itemize}
    \item What is the relative performance of different multi-market bidding strategies?
    \item To what extent does market liquidity, modeled via bid-ask spreads in equations (\ref{bid-price}) and (\ref{ask-price}), affect the performance of the rolling intrinsic strategy?
    \item How robust is the rolling intrinsic method to trading bucket size?
    \item How sensitive are the strategy outcomes to variations in battery system parameters, such as maximum cycle limits, charge and discharge power, and the initial and final state of charge (SOC)?
\end{itemize}

We evaluate the following strategy configurations:

\begin{itemize}
\item \textbf{DA}: Battery dispatch optimized using day-ahead auction spot prices (EPEX auction at 12:00 CET).
\item \textbf{ID\_AUCT}: Optimization based on the EPEX IDA1 intraday auction prices (auction at 15:00 CET).
\item \textbf{ID\_AEP}: Optimization using the ID AEP price index, as published on Netztransparenz.de \cite{Netztransparenz}.
\item \textbf{ID1}, \textbf{ID3}, \textbf{IDFULL}. Optimization based on the respective EPEX intraday trading Indices.
\item \textbf{ID\_ROLL}: Rolling intrinsic optimization using continuous bid and ask prices as defined in equations (\ref{bid-price}) and (\ref{ask-price}), following the procedure outlined in Algorithm \ref{ri-algorithm}.
\item \textbf{X$|$Y}: Hybrid strategy where initial dispatch is optimized using market X, and subsequent redispatch is conducted using market Y. For example, DA$|$ID\_AUCT refers to an initial optimization based on DA prices, followed by redispatch using ID\_AUCT prices.
\end{itemize}

It should be emphasised that the optimisations based on the indices ID\_AEP, ID1, ID3, and IDFULL are reported only as benchmarking tools. These indices cannot be traded directly, as they are defined ex post and therefore lack practical applicability in real-world trading. Nonetheless, we include them in the analysis because they are frequently used in the literature as proxies for intraday market prices, and their comparison provides useful insights into potential value differences. In contrast, all other strategies presented in this study correspond to implementable trading approaches that could realistically be executed in practice with reasonable effort.

For the day-ahead strategy, a price forecast is required. This forecast can be generated via an internal fundamental model or obtained from commercial providers. Based on our experience, EXAA day-ahead prices do also serve as a reliable proxy. The resulting optimal schedule can be submitted to the market as a Loop Block order \cite{EPEX2025}.

Similarly, for intraday auction-based strategies, forecasted price data is necessary. These forecasts may be purchased, or alternatively, mid-prices from the order book  can be used as reasonable estimates for auction outcomes when available.

\begin{table}\footnotesize
    \centering
    \begin{tabular}{llrrrrr}
         & bidding strategy & mean & median & std & min & max \\
        \hline
         \multirow{3}{*}{single-market} &DA & 228.75 & 216.05 & 149.20 & 15.52 & 1485.61 \\
         & ID\_AUCT & 287.09 & 255.66 & 255.93 & 61.08 & 3924.22 \\
         & ID\_ROLL & 296.60 & 264.20 & 185.73 & 63.72 & 1982.93 \\
        \hline
        
        \multirow{4}{*}{multi-market} & DA$|$ID\_AUCT & 285.30 & 267.20 & 173.94 & 47.91 & 1633.71 \\
         & DA$|$ID\_ROLL & 315.83 & 295.99 & 185.43 & 61.24 & 2027.17 \\
        & ID\_AUCT$|$ID\_ROLL & 340.93 & 306.04 & 309.50 & 94.47 & 4957.51 \\
        & DA$|$ID\_AUCT$|$ID\_ROLL & 339.15 & 312.17 & 215.18 & 81.30 & 2297.83 \\
        \hline
        \hline
        \multirow{4}{*}{indices} & ID1 & 337.75 & 261.28 & 401.74 & 31.31 & 5631.55 \\
        & ID3 & 293.48 & 246.09 & 264.43 & 30.06 & 3815.98 \\
        & IDFULL & 301.84 & 250.38 & 284.73 & 29.20 & 3707.41 \\
        & ID\_AEP & 453.10 & 299.95 & 693.66 & 31.64 & 7556.81 \\
    \end{tabular}
\caption{Statistical measures of profit (in \pnlunit) for period  14/6/2024 to 1/7/2025 for 2h-battery (see Table \ref{tab:battery_confiurations}).}\label{tab:mm-bidding-results}
\end{table}
\begin{table}\footnotesize
\centering
    \begin{tabular}{lrrrrr}
    products traded & mean & median & std & min & max \\
    \hline
    15 min products & 296.60 & 264.20 & 185.73 & 63.72 & 1982.93 \\
    1h products & 240.43 & 219.20 & 157.06 & 52.74 & 1570.67 \\
    \end{tabular}
\caption{Profit statistics (in \pnlunit)  for period  14/6/2024 to 1/7/2025 and a 2h-battery (see Table \ref{tab:battery_confiurations}) using the rolling intrinsic strategy in intraday continuous trading with five minute frequency. Here, either quarter hour products or hour products are traded.}\label{tab:1h_vs_15min_rolling_intrinsic}
\end{table}

\begin{figure}
    \centering
    \includegraphics[width=0.49\textwidth]{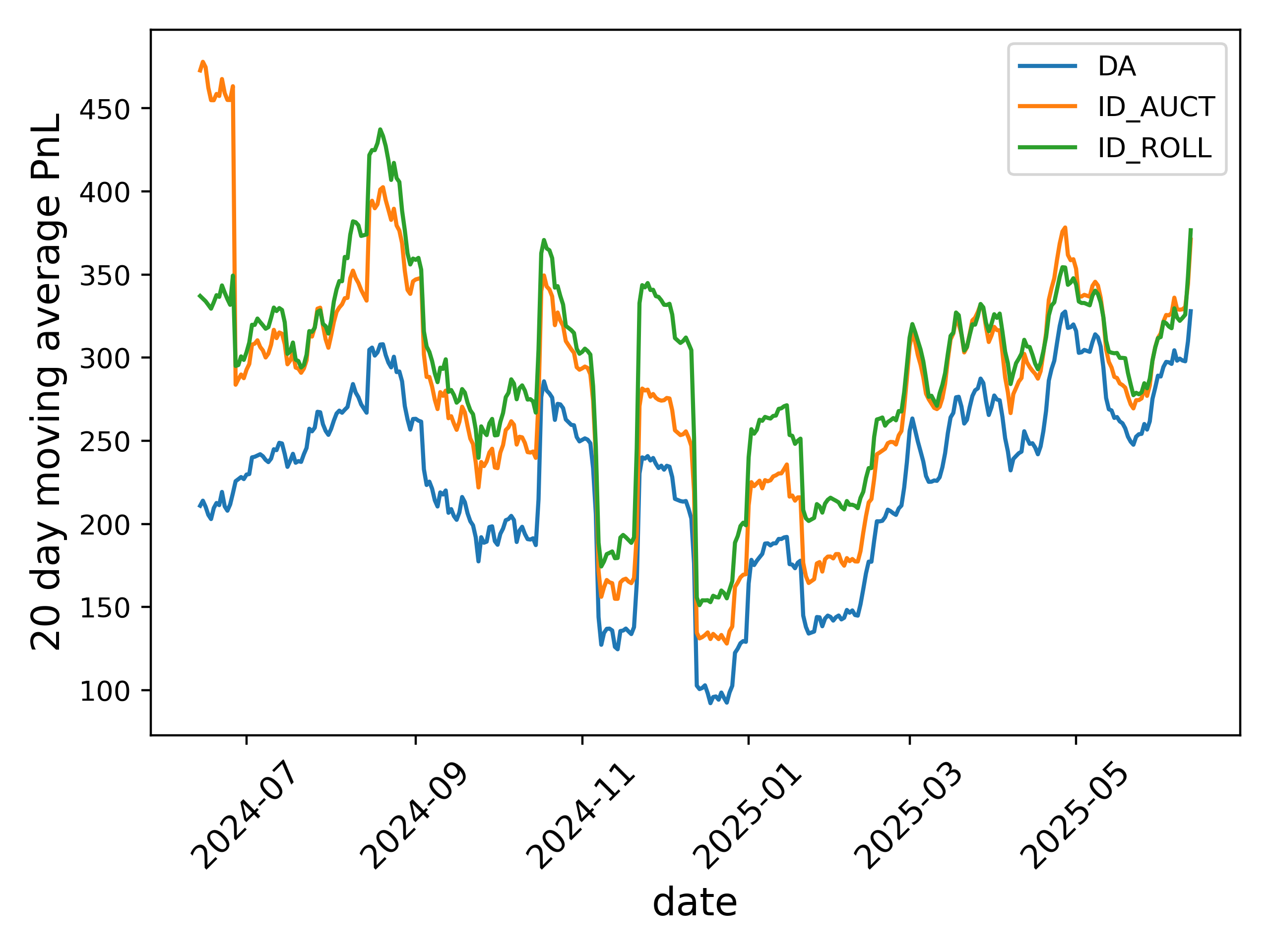}
    \includegraphics[width=0.49\textwidth]{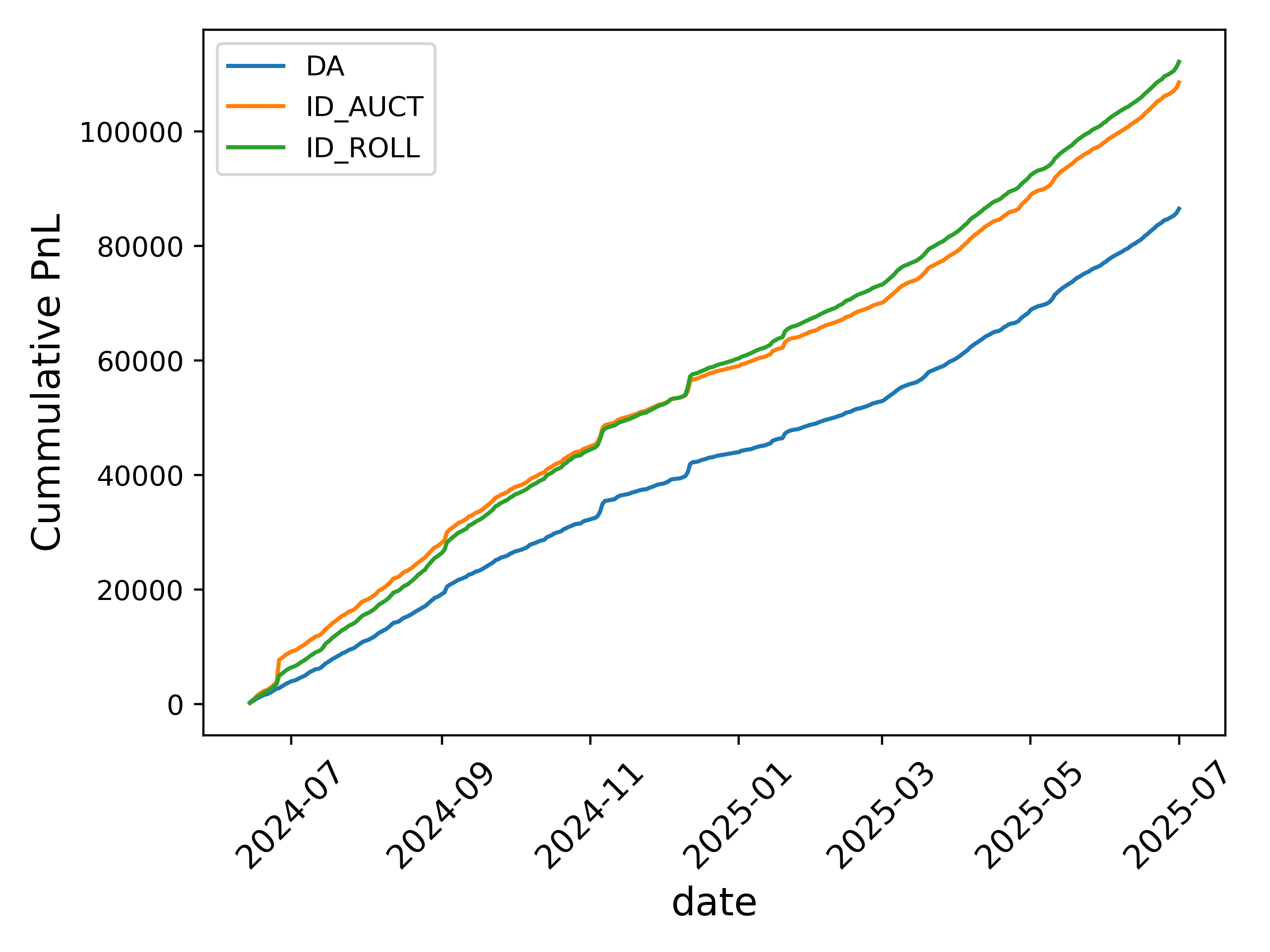}
    \caption{20 day moving average of profit (in \pnlunit) for the single-market bidding strategies for a 2-h battery (left) and the series of the cumulative sum of the profits.}
    \label{fig:single-market-bidding}
\end{figure}
\begin{figure}
    \centering
    \includegraphics[width=0.32\textwidth]{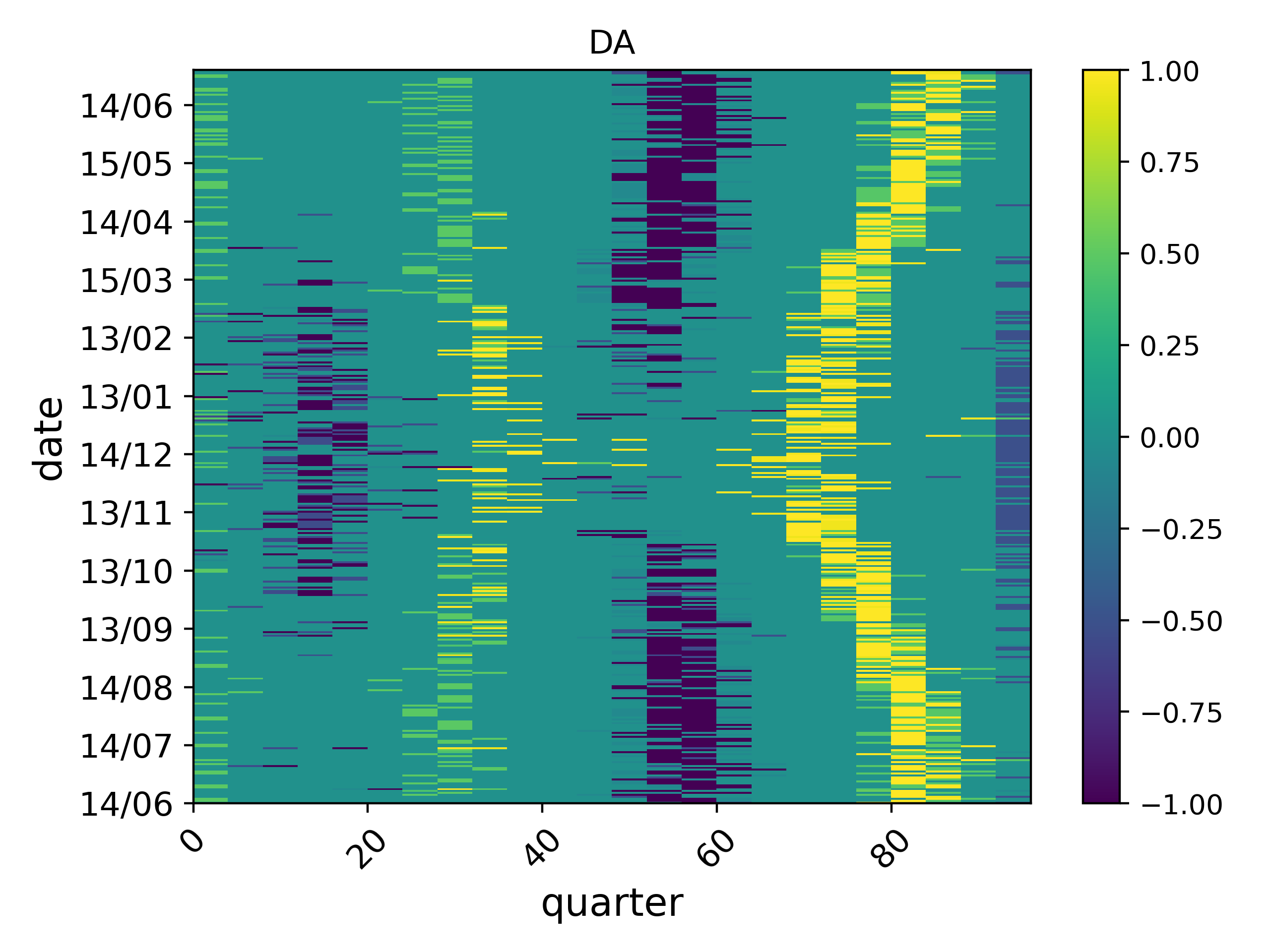}
    \includegraphics[width=0.32\textwidth]{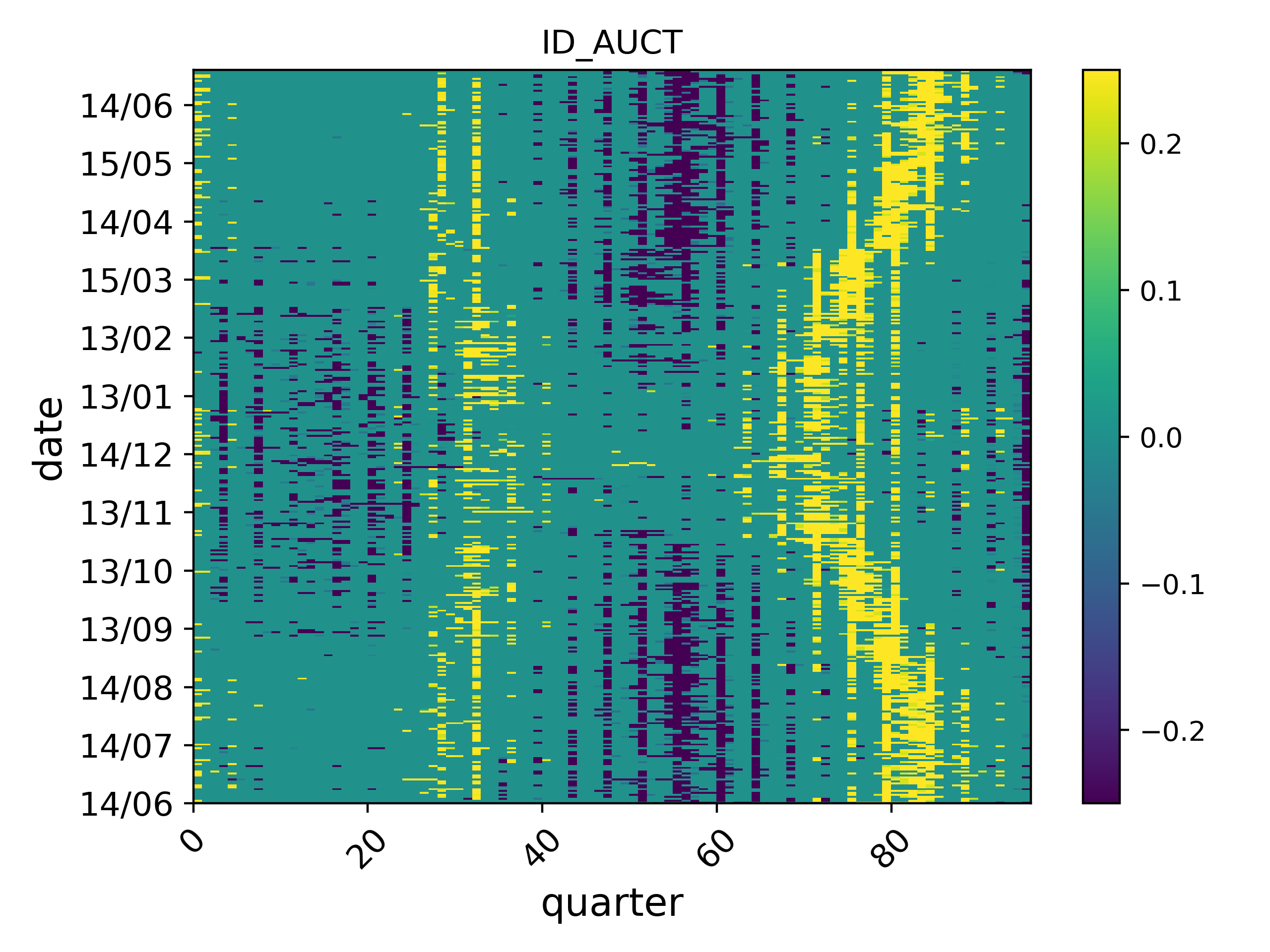}
    \includegraphics[width=0.32\textwidth]{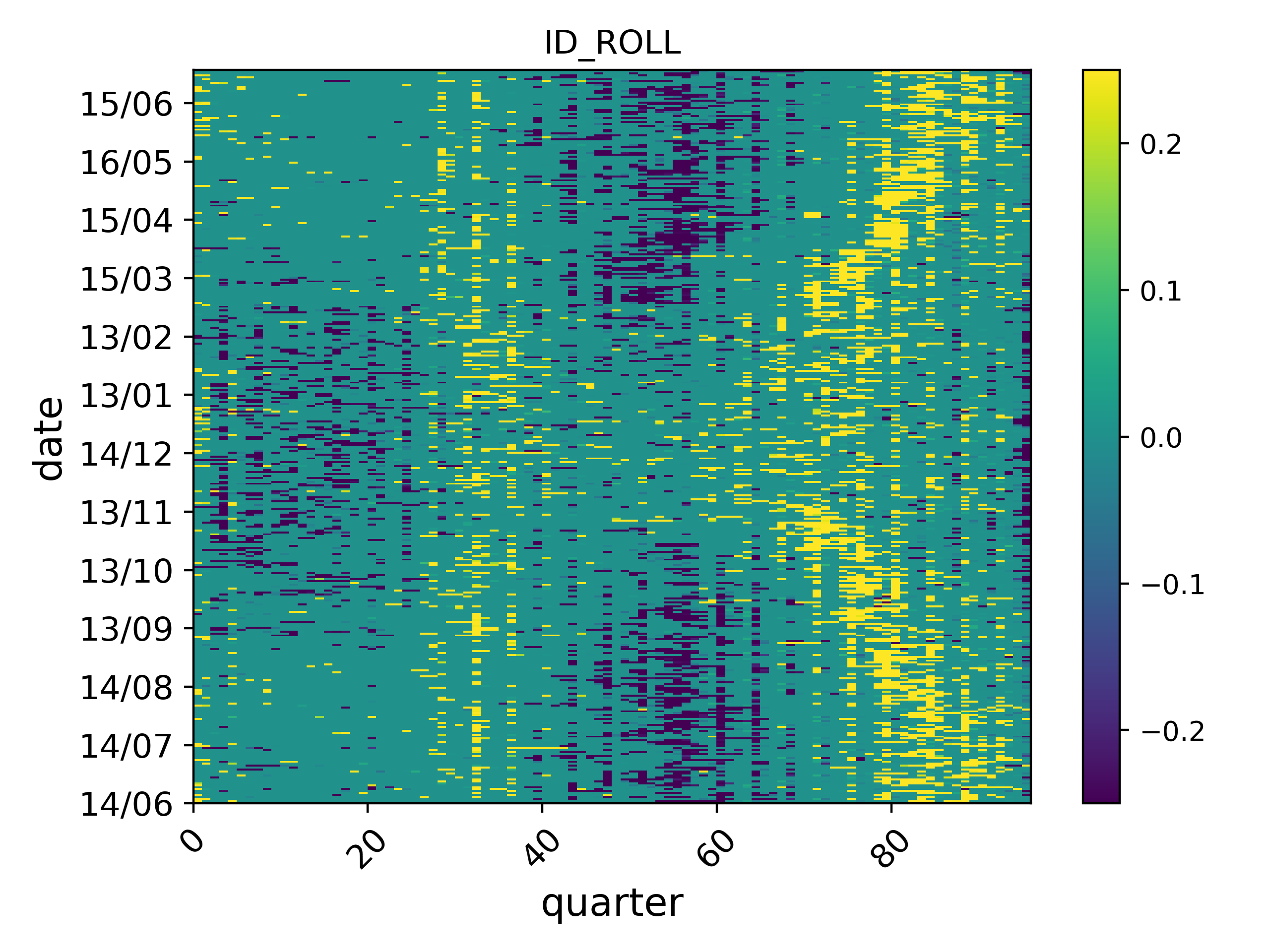}\\
    \includegraphics[width=0.32\textwidth]{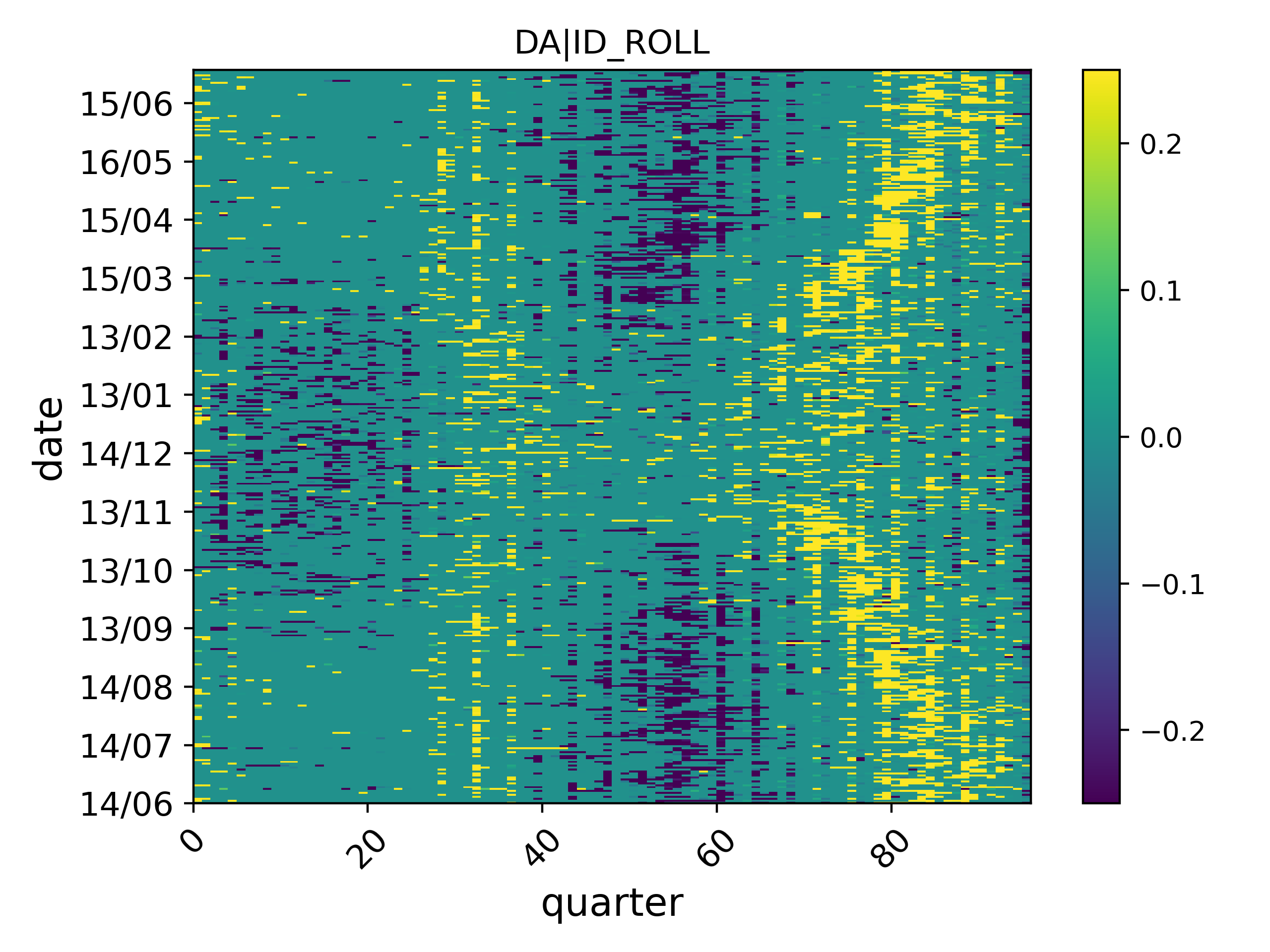}
    \includegraphics[width=0.32\textwidth]{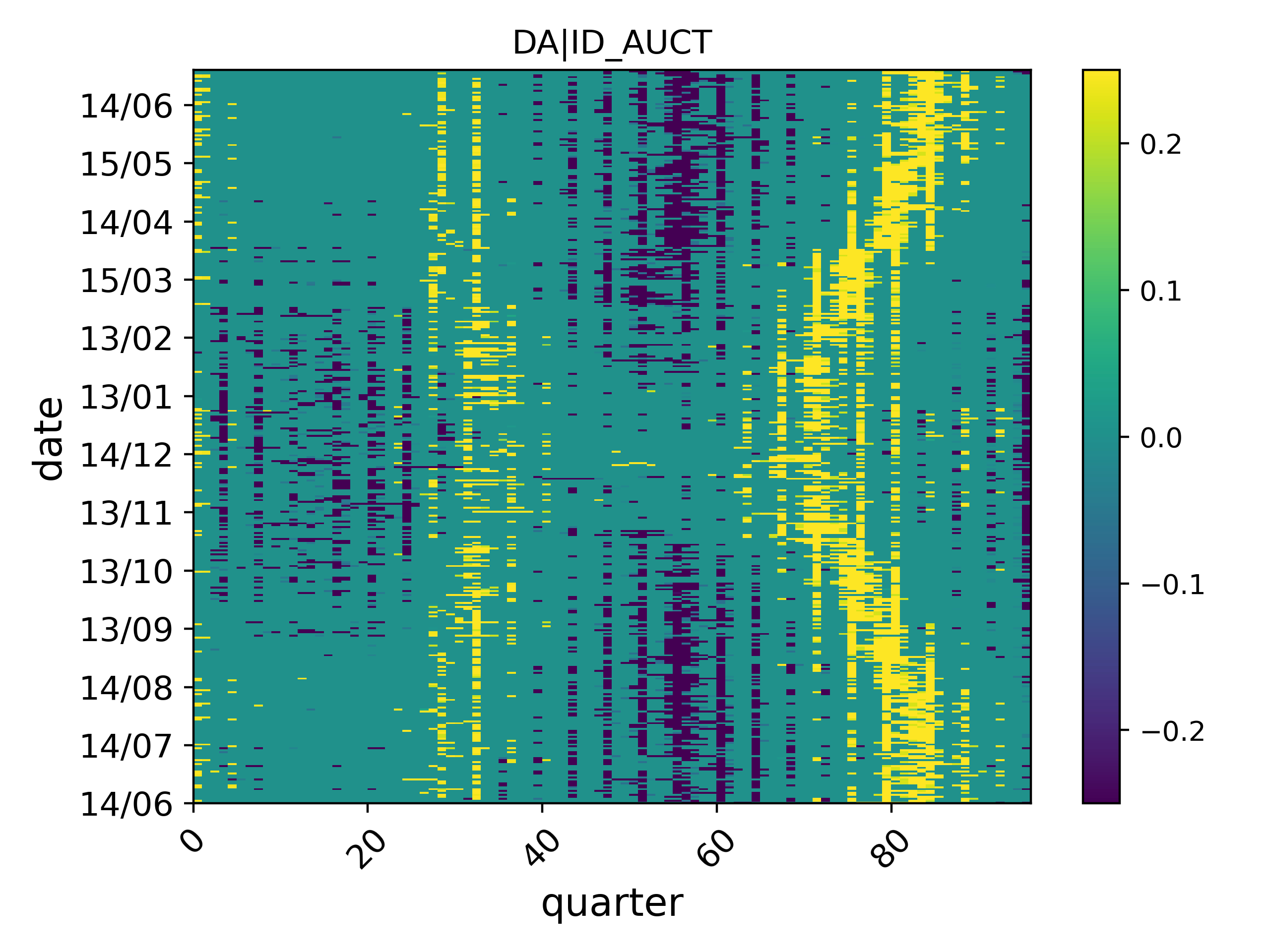}
    \includegraphics[width=0.32\textwidth]{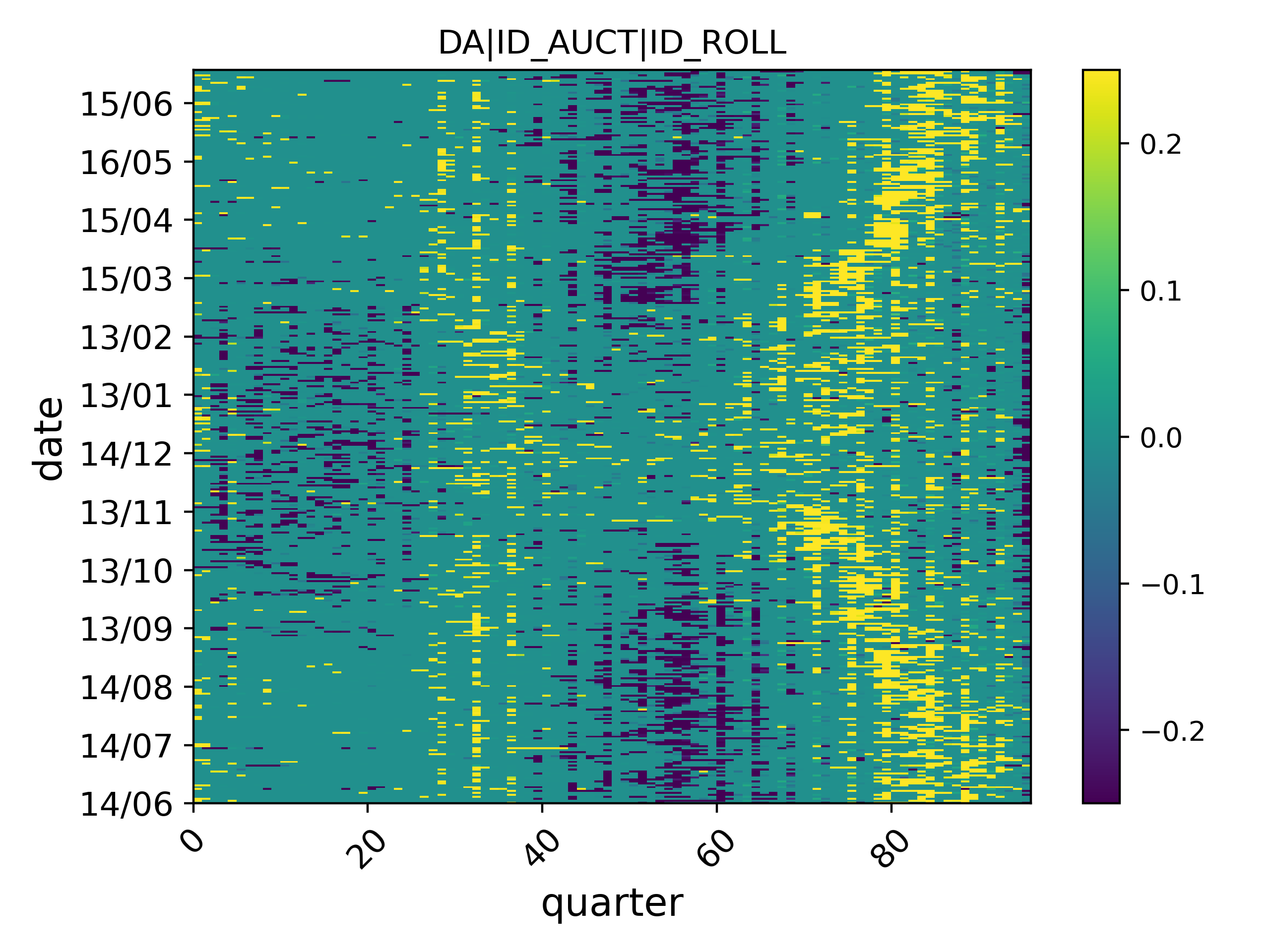}
    \caption{History of dispatch for different market bidding strategies (negative values indicate charging, positive discharging).}
    \label{fig:market-bidding-dispatch}
\end{figure}
\begin{figure}
    \centering
    \includegraphics[width=0.32\textwidth]{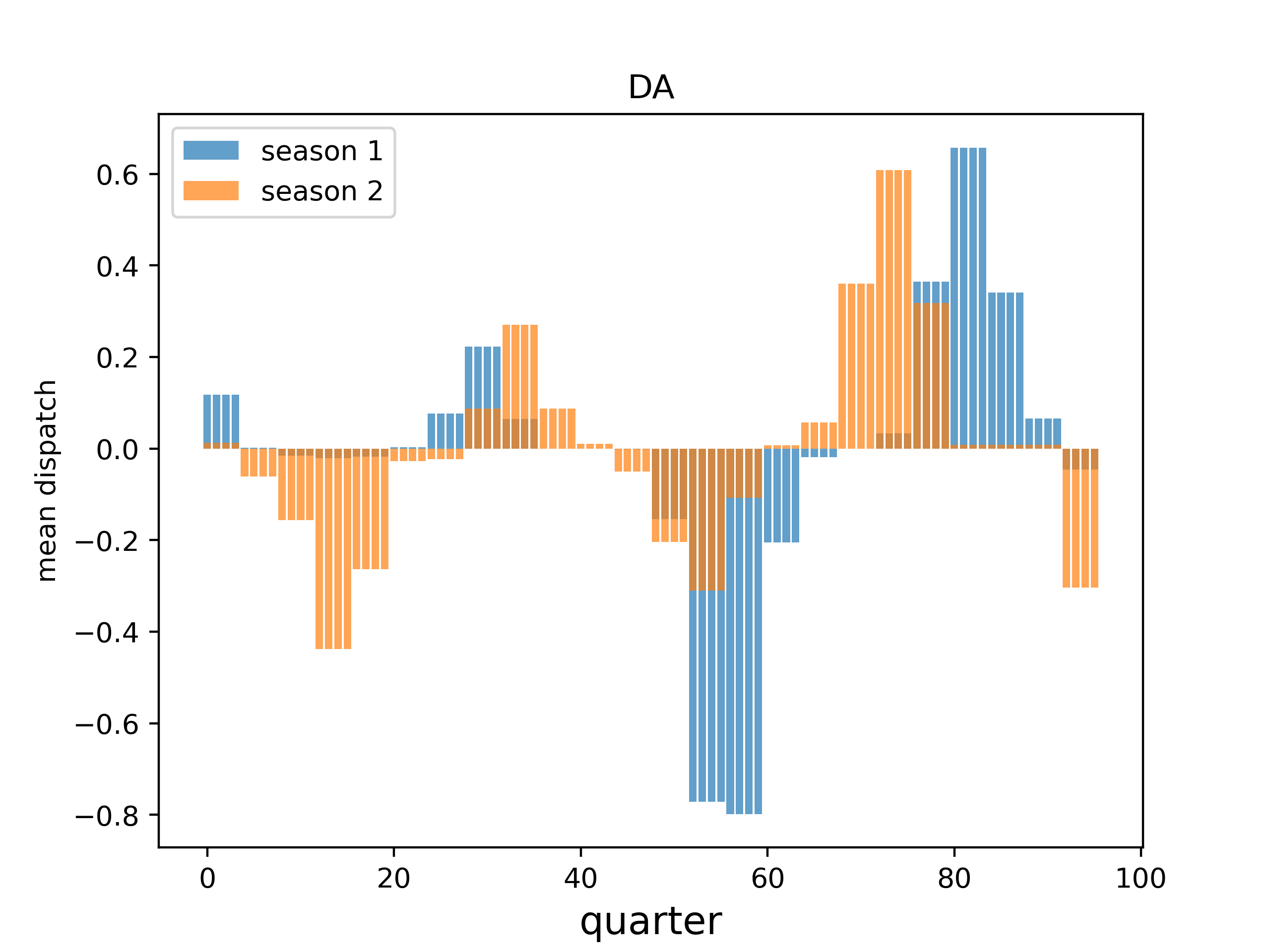}
    \includegraphics[width=0.32\textwidth]{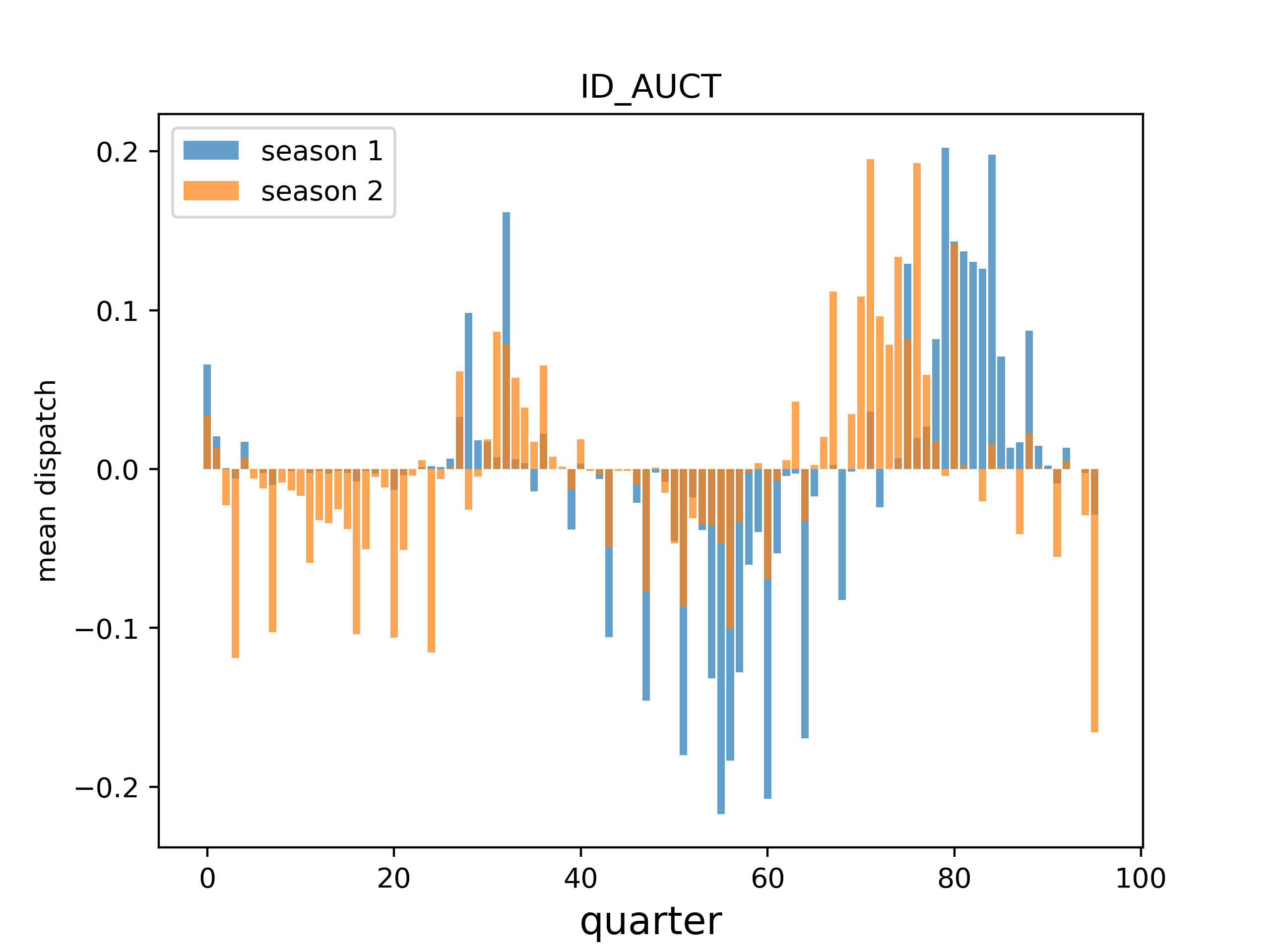}
    \includegraphics[width=0.32\textwidth]{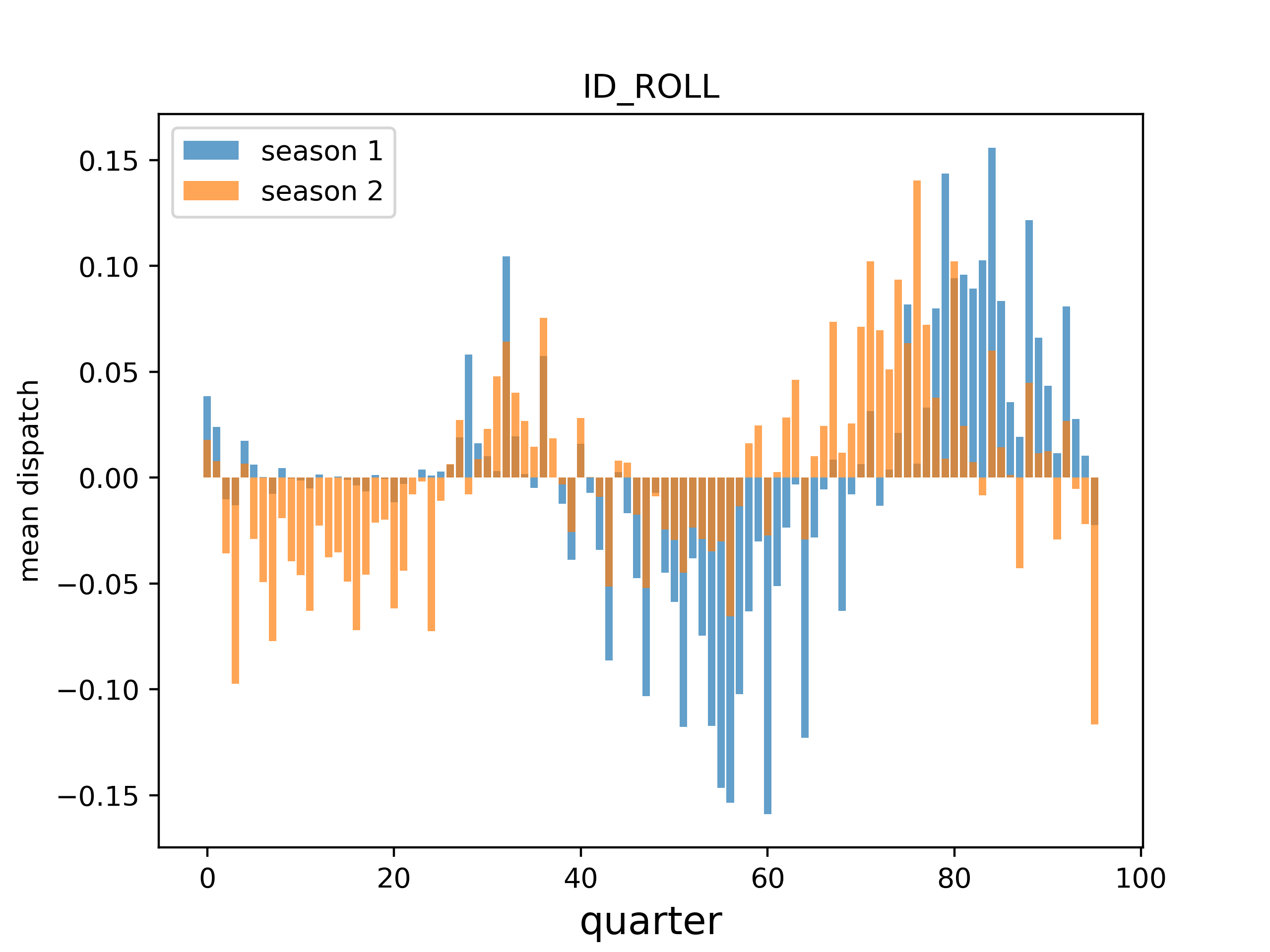}
    \caption{Mean of dispatches for different strategies over the history divided into two seasons: Season 1 includes April to September and season 2 the other months.} \label{fig:market-bidding-mean-dispatch}
\end{figure}
\paragraph{Multi-market bidding:} 
Table~\ref{tab:mm-bidding-results} presents summary statistics of the realized profits for the period from June 14, 2024, to July 1, 2025, across the different bidding strategies. The results indicate that multi-market bidding strategies generally yield higher profits compared to single-market approaches. An exception is observed in the case of the day-ahead bidding strategy followed by redispatching in the intraday auction, which achieves profit levels comparable to those obtained by directly dispatching based solely on the intraday auction prices.

Consistent with the findings of \cite{Loehndorf2023} and in line with the empirical results reported in \cite{Miskiw2025}, multi-market bidding strategies that include participation in the day-ahead market tend to perform similarly or even less favorably compared to strategies that focus solely on intraday auction trading, followed by redispatching via the rolling intrinsic approach in the continuous intraday market.

This effect is even more pronounced in the case of single-market bidding, as illustrated in the right panel of Figure~\ref{fig:single-market-bidding}, which displays the cumulative profit-and-loss (PnL) trajectories. While the results in Table~\ref{tab:mm-bidding-results} indicate that the standard deviation of profits is lower for day-ahead bidding relative to intraday or rolling intrinsic strategies, this lower variability does not reflect improved risk-adjusted performance. Rather, it is a consequence of a reduced range of attainable profits, particularly on the upside, which is not advantageous from an economic standpoint. This observation is further supported by the left panel of Figure~\ref{fig:single-market-bidding}, which plots the 20-day moving average of profits and highlights the limited profit potential associated with exclusive reliance on day-ahead market participation.

As expected, all markets display a comparable dispatch pattern, as shown in Figure \ref{fig:market-bidding-dispatch}. Negative values represent charging activity, while positive values indicate discharging. In winter, the timing of charging shifts from midday to the early morning hours, whereas discharging moves from the evening to the afternoon and late morning. While dispatch strategies based on the rolling intrinsic approach exhibit slightly higher volatility, they retain the same fundamental characteristics.
These seasonal patterns become even more apparent in Figure~\ref{fig:market-bidding-mean-dispatch}, which shows the mean dispatch profiles averaged over two periods: April to September (spring/summer) and October to March (autumn/winter).

\paragraph{Intraday trading 15min vs 1h products:} As outlined in Section \ref{sec:market-statistics}, intraday markets also feature hourly products, which generally exhibit higher trading volumes than the corresponding 15-minute products. This raises the question of whether hourly products could present a viable alternative for battery arbitrage compared to 15-minute products. Table \ref{tab:1h_vs_15min_rolling_intrinsic} reports results from the rolling intrinsic strategy applied to both product types. Consistent with the prevailing view, that the greater flexibility of 15-minute products enhances arbitrage opportunities, our findings show that trading 15-minute products yields, on average, nearly 20\% higher returns than hourly contracts.

\paragraph{Effect of bid-ask spreads and trading frequency:} 
\begin{table}
\footnotesize
\centering
    \begin{tabular}{lrrrrr}
     & mean & median & std & min & max \\
    \hline
    base (5 min)  & 296.60 & 264.20 & 185.73 & 63.72 & 1982.93 \\
    50\% quantile & 323.37 & 286.25 & 213.50 & 78.24 & 2478.72 \\
    30 min        & 255.67 & 221.67 & 168.38 & 7.95 & 1691.49 \\
    \end{tabular}
    \caption{Profit statistics (in \pnlunit)  for period 14/6/2024 to 1/7/2025 of ID\_ROLL for 2h-battery using the default settings (5 minute trading buckets and 20\% quantile to create bid-ask prices) in comparison to median value (bid equals ask price) and 30 minute trading buckets. }\label{tab:bid-ask-spreads}
\end{table}

The previous results showed that including intraday trading using the rolling intrinsic approach after initial auction marketing improves profitability. Here, the construction of bid-ask prices as described in (\ref{bid-price}) and (\ref{ask-price}) may have a significant influence on the resulting profit. To investigate the effects of the chosen quantile within the construction as well as the trading frequency, we present results for ID\_ROLL in Table \ref{tab:bid-ask-spreads}:
%

%A decrease of intraday trading frequency that by %construction also leads to an increase in bid-ask spreads, leads to a %decrease of approximately 14\% to a mean profit of 323 EUR compared to the %base scenario with 297 EUR. Eliminating the bid-ask spread  by using a 50\% %quantile increases the profit by around 9\% up to 323 EUR.
%
A decrease of intraday trading frequency from 5 to 30~min intervals, which by construction also leads to an increase in bid-ask spreads, leads to a decrease of approximately 14\% to a mean profit of 256~\pnlunit compared to the base scenario with 297~\pnlunit. Eliminating the bid-ask spread by using a 50\% quantile increases the profit by around 9\% up to 323~\pnlunit.

\paragraph{Value of Charge and Discharge Capacities:} 
\begin{table}
\footnotesize
\centering
\begin{tabular}{lccc}
  battery & DA & DA$|$ID\_AUCT & DA$|$ID\_AUCT$|$ID\_ROLL \\
    \hline
    1h & 241.25 & 318.88 & 401.44 \\
    2h & 228.75 & 285.30 & 339.15 \\
    4h & 198.31 & 234.07 & 274.54 \\
\end{tabular}
\caption{Mean profit for batteries (in \pnlunit) with different C-rates for period 14/6/2024 to 1/7/2025 and different strategies.}\label{tab:c-rate-results}
\end{table}
In this section, we analyze the variation in economic value associated with different charge and discharge capacities, commonly expressed in terms of the C-rate. Specifically, we examine the performance of batteries with one-hour, two-hour, and four-hour durations, as specified in Table \ref{tab:battery_confiurations}. For investment decisions, it is essential to estimate the potential profits achievable at different C-rates, particularly given that battery capital costs are typically dependent on this parameter. Table \ref{tab:c-rate-results} presents the mean profits obtained over the period from 14 June 2024 to 1 July 2025, based on three operational strategies:  day-ahead optimization (DA), day-ahead optimization with redispatch (DA$|$ID\_AUCT) via the intraday auction, and  redispatch via the intraday auction followed by continuous trading using a rolling intrinsic approach (DA$|$ID\_AUCT$|$ID\_ROLL). We see that the increase in value between the 1-hour and 2-hour battery for the observed period is far below the theoretical upper bound of 200\%  and just 5\% for the day-ahead market, 12\% between bidding on the day-ahead market and redispatching on the intraday auction and 18\% adding a rolling intrinsic strategy at the end. 
%\paragraph{Effect of initial and final SoC:} All strategies we investigated so far had the same half-full state of charge at the beginning and end of the trading day. Note that since \marginpar{Hier weiterschreiben!!!!}%%%%%%%%%%%%%%
%%%%%%%%%%%%%%%%%%%%%%%%%%%%%%%%%%%%%%%%%%%%%%%%%%%%%%%%%%%%%%
%%%%%%%%%%%%%%%%%%%%%%%%%%%%%%%%%%%%%%%%%%%%%%%%%%%%%%%%%%%%%
%%%%%%%%%%%%%%%%%%%%%%%%%%%%%%%%%%%%%%%%%%%%%%%%%%%%%%%%%%%%
\paragraph{Effect of cycle limitations:} 
\begin{table}
\centering
    \begin{tabular}{lrrrrr}
    label & mean & median & std & min & max \\
    \hline
    1 daily cycle & 340.93 & 306.04 & 309.50 & 94.47 & 4957.51 \\
    2 daily cycles & 466.16 & 411.27 & 387.35 & 127.49 & 6267.92 \\
    3 daily cycles & 530.13 & 462.83 & 427.35 & 168.92 & 6806.51 \\
    4 daily cycles & 559.13 & 489.53 & 439.79 & 164.48 & 6889.85 \\
    \end{tabular}
\caption{Profit statistics (in \pnlunit) for a 2-hour battery with market bidding at the intraday auction followed by a rolling intrinsic strategy. }
    \label{tab:max_cycle_constraints}
\end{table}
\begin{figure}
    \centering
    \includegraphics[width=0.49\textwidth]{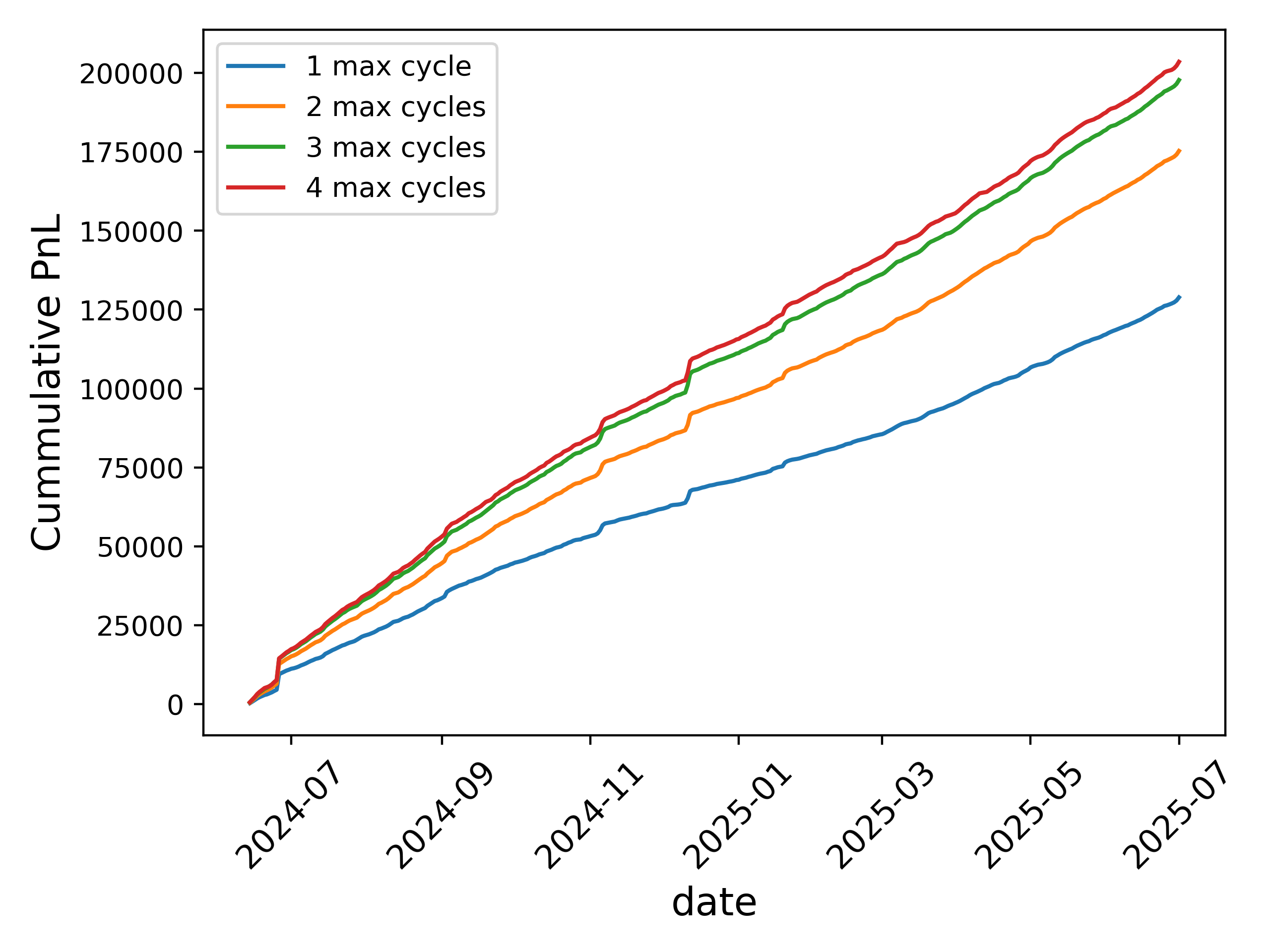}
    \includegraphics[width=0.49\textwidth]{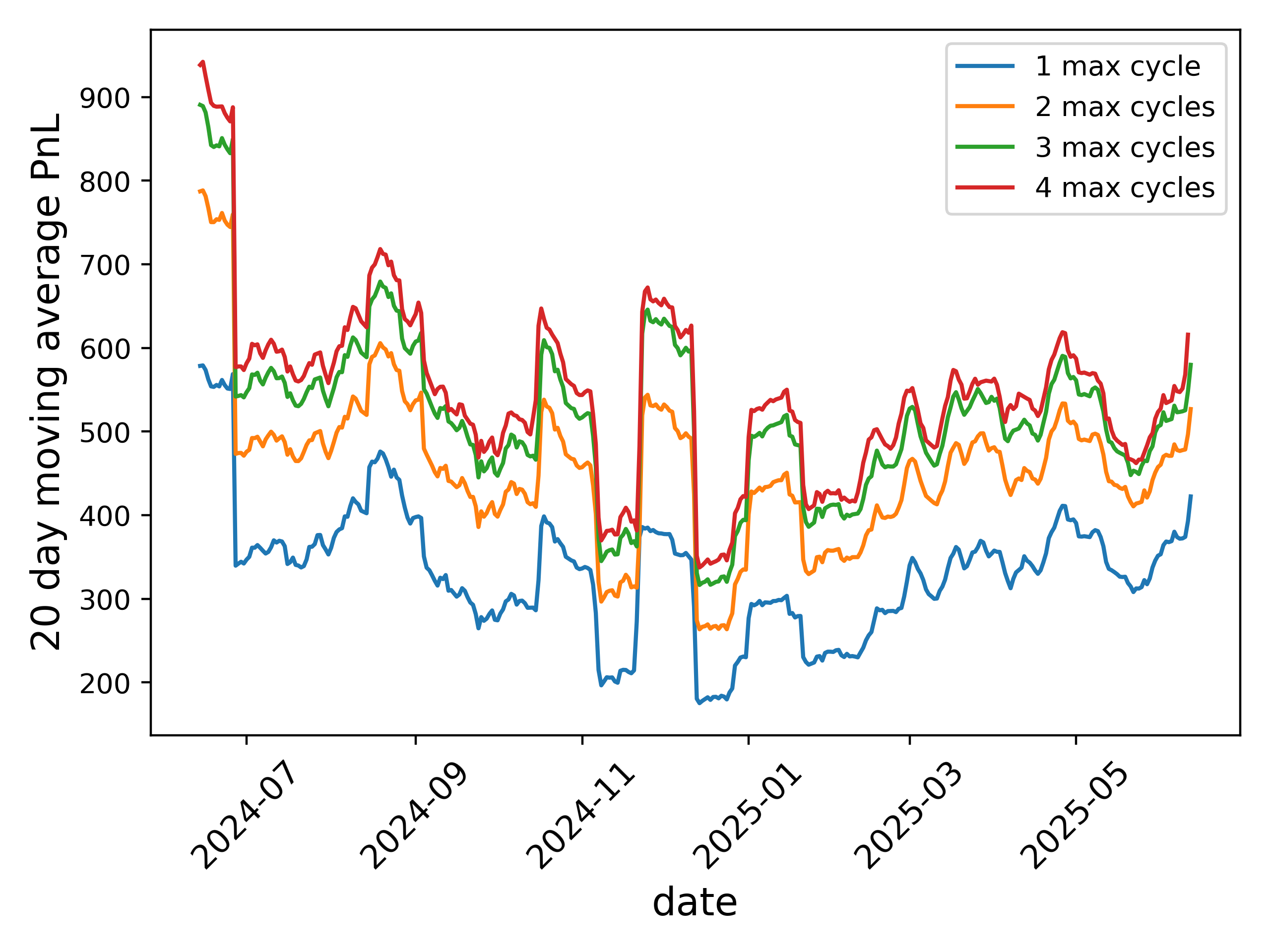}
    \caption{Cummulative sum of profits for different maximum cycle constraints (left) and 20 day moving average of profits (in \pnlunit) (right).}
    \label{fig:max_cycles_profit}
\end{figure}
\begin{table}[]
\footnotesize
    \centering
    \begin{tabular}{lrrrrr}
     & mean & median & std & min & max \\
    \hline
    0\% & 340.93 & 306.04 & 309.50 & 94.47 & 4957.51 \\
    5\% of days & 351.36 & 305.67 & 400.40 & 0.00 & 6267.92 \\
    10\% of days & 352.15 & 305.67 & 409.51 & 0.00 & 6267.92 \\
    15\% of days & 351.47 & 307.25 & 417.98 & 0.00 & 6267.92 \\
    20\% of days & 349.50 & 314.24 & 425.10 & 0.00 & 6267.92 \\
    25\% of days & 346.17 & 315.94 & 433.23 & 0.00 & 6267.92 \\
\end{tabular}
\caption{Statistics of profits (in \pnlunit) for different strategies that suspend operations on less profitable days. The percentage indicates the share of days on which operations may be suspended.}
    \label{tab:max_cycle_strategy}
\end{table}
In the baseline setting, battery operation was restricted to one full load cycle per day. We now relax this assumption and assess the effect of allowing for multiple cycles. Table \ref{tab:max_cycle_constraints} reports profit statistics for up to four daily cycles of the 2h-battery configuration (see Table \ref{tab:battery_confiurations}). The results show diminishing marginal returns: the increase from one to two cycles adds about 130~\pnlunit on average, while expanding from two to four cycles yields only an additional 93~\pnlunit. Figure \ref{fig:max_cycles_profit} illustrates this trend with cumulative profits and a 20-day moving average of daily profits.

Daily cycle limits are a modeling simplification, as battery warranties typically specify annual cycle constraints. Daily constraints are often used either to reduce optimization complexity or to align with service provider contracts. A flexible yearly allocation could, in principle, improve profitability by concentrating operation on high-profit days and reducing activity on low-profit days.

Closer inspection, however, shows limited potential. A single cycle generates on average 340~\pnlunit, whereas the incremental profit from a second cycle is only 130~\pnlunit. To test the impact of flexible allocation, we apply an ex-post strategy: for a given share of days, operation is suspended on the least profitable days and replaced with two cycles on the days where the incremental benefit is highest. Results in Table \ref{tab:max_cycle_strategy} show that under this optimized allocation, profits improve by at most 3.2\% compared to the baseline of one cycle per day. This shows that operating with more than one cycle per day--increasing annual cycles--can generate non-negligible additional revenues. Consequently, operators must balance the short-term, relatively secure gains from additional cycling against the long-term costs, which may include accelerated battery degradation, warranty violations, or reduced operational flexibility later in the year. This trade-off underscores the importance of explicitly accounting for warranty structures and long-term asset value when designing dispatch strategies.

%These findings show, under the observed market conditions, the potential to increase profits through more sophisticated heuristics for managing cycle constraints. Nevertheless, Table \ref{tab:max_cycle_constraints} shows 

\section{Conclusion}
This paper has examined the profitability of different bidding strategies for Battery Energy Storage Systems (BESS) in the Central European wholesale power markets, with a particular focus on the interplay between day-ahead and intraday markets. Our analysis builds on the rolling intrinsic approach, which proved to be a robust method for intraday trading by dynamically capturing short-term price fluctuations. To better approximate market realities, liquidity constraints were explicitly represented through bid--ask spreads, highlighting their significant impact on achievable revenues.

The results demonstrate that multi-market bidding strategies consistently outperform single-market participation, regardless of whether this is limited to the day-ahead auction, the intraday auctions, or continuous intraday trading. While pure intraday, i.e. first the intraday auction followed by continuous intraday trading, tend to achieve the highest returns, integrating the day-ahead market into multi-market strategies slightly reduces profitability. Nevertheless, from a risk perspective, participation in both markets remains attractive: bidding in the day-ahead auction as well as the intraday auction increases the likelihood of successfully securing profitable trades, thereby reducing the risk of non-execution that arises when relying on a single market alone. These findings are in line with recent literature showing that coordinated or combined market strategies can improve the robustness of storage operation under uncertainty \cite{Miskiw2025, Loehndorf2023}.

Finally, the analysis of maximum cycle limits reveals further potential for more sophisticated bidding strategies. In particular, strategies that relax strict daily cycling constraints while ensuring compliance with annual throughput restrictions could unlock additional value. This suggests that the design of intertemporal constraints plays a crucial role in capturing the full economic potential of storage assets and deserves more attention in future research.

Building on these insights, several directions for future research emerge. First, extending the rolling intrinsic approach with stochastic forecasting methods could better capture price uncertainties and improve decision-making under volatile market conditions. Here, due to the high dimensionality of the problem, a deep hedging approach involving neural networks \cite{buehler_2019} which has been successfully applied in the energy context for green PPAs \cite{BieglerKoenig2025} may be an interesting direction.
Second, exploring market coupling beyond energy-only products, such as integrating reserve or balancing markets, may provide additional revenue streams and risk-hedging opportunities for BESS operators. Third, further work should investigate long-term operational constraints, particularly strategies that coordinate daily and yearly cycling requirements in a unified framework. 

\bibliography{literature}

\end{document}